\documentclass[aps,prc,a4paper,nofootinbib,preprintnumbers,superscriptaddress,showpacs,showkeys,twocolumn]{revtex4}

\usepackage{amsmath}
\usepackage{amssymb}
\usepackage{bm}
\usepackage{graphicx}
\usepackage{color}
\usepackage[english]{babel}

\newcommand{\be}{\begin{equation}}
\newcommand{\ee}{\end{equation}}
\newcommand{\ba}{\begin{eqnarray}}
\newcommand{\ea}{\end{eqnarray}}

\begin{document}

\title{Thermodynamic Geometry of the Quark-Meson Model}

\author{Bonan Zhang}
\affiliation{School of Nuclear Science and Technology, Lanzhou University, 222 South Tianshui Road, Lanzhou 730000, China}
\author{Shen-Song Wan}
\affiliation{School of Nuclear Science and Technology, Lanzhou University, 222 South Tianshui Road, Lanzhou 730000, China}
\author{Marco Ruggieri}\email{ruggieri@lzu.edu.cn}
\affiliation{School of Nuclear Science and Technology, Lanzhou University, 222 South Tianshui Road, Lanzhou 730000, China}


\begin{abstract}
We study the thermodynamic geometry of the Quark-Meson model, focusing on the curvature, $R$, around the chiral crossover at
finite temperature and baryon chemical potential. We find a peculiar behavior of $R$ in the crossover region,
in which the sign changes and a local maximum develops; in particular, the height of the peak of $R$ in the crossover region becomes large in proximity of the critical
endpoint and diverges at the critical endpoint.
The appearance of a pronounced peak of $R$ close to the critical endpoint supports the idea that $R$ grows with the correlation volume around the phase transition.
We also analyze the mixed fluctuations of energy and baryon number, $\langle\Delta U\Delta N\rangle$,
which grow up substantially in proximity of the critical endpoint: in the language of thermodynamic geometry these fluctuations
are responsible for the vanishing of the determinant of the metric, which results in thermodynamic instability
and are thus related to the appearance of the second order phase transition at the critical endpoint.  
\end{abstract}



\maketitle

\section{Introduction}

An interesting idea of Statistical Mechanics is that of a metric in the variety spanned by the thermodynamic variables: this is related to the theory of
fluctuations among equilibrium states and lead to the concept of thermodynamic geometry and thermodynamic curvature  \cite{Weinhold:1975get,Weinhold:1975gtii,Ruppeiner:1979trg,Ruppeiner:1981agt,Ruppeiner:1983ntp,Ruppeiner:1983tcf,
Ruppeiner:1985cgt,Ruppeiner:1985tcv,Ruppeiner:1986tcv,Ruppeiner:1990tig,Ruppeiner:1990tof,Ruppeiner:1991rgc,
Ruppeiner:1993afg,Ruppeiner:1995rgf,Ruppeiner:1998rgc,Ruppeiner:2005rgr,Ruppeiner:2008tcb,Ruppeiner:2010tci,Ruppeiner:2012tct,
Ruppeiner:2012tpw,Ruppeiner:2013tfb,Ruppeiner:2013tar,Ruppeiner:2014utg,Ruppeiner:2015trf,Ruppeiner:2015tsf,Ruppeiner:2015tsw,
Ruppeiner:2016smt,Ruppeiner:2017svc,Wei:2013ctb,Janyszek:1989rtm,Janyszek:1,Janyszek:2,
Sahay:2017gcb,Castorina:2018ayy,Mirza:2009nta,Mirza:2008rag,Castorina:2018gsx,Castorina:2019jzw,
Ruppeiner:maybe,covariant:evolution,geometrical:aspects,crooks:measuring,bellucci:PA}.
For example, in the grand-canonical ensemble the equilibrium state is specified as long as the intensive independent variables like
temperature, chemical potential and others are fixed, and physical quantities like energy and particle number fluctuate
with probability given by the Gibbs ensemble. Considering the pair of intensive variables $(\beta^1,\beta^2)$
the probability of a fluctuation from $S_1=(\beta^1,\beta^2)$ to $S_2=(\beta^1 + d\beta^1,\beta^2+d\beta^2)$ is proportional to
\begin{equation}
\sqrt{g}\exp\left(-\frac{1}{2}g_{ij}d\beta^i d\beta^j\right),
\end{equation}
where $g_{ij}\equiv\partial^2\log{\cal Z}/\partial\beta^i\partial\beta^j$ is called the thermodynamic metric tensor,
$g=\mathrm{det}(g_{ij})$ is the determinant of $g_{ij}$
and ${\cal Z}$ is the grandcanonical partition function. It is therefore natural to define the line element
$d\ell^2 = g_{ij}d\beta^i d\beta^j$ which measures effectively a distance between $S_1$ and $S_2$,
in the sense that a large $d\ell^2$ corresponds to a small probability of a fluctuation from $S_1$ to $S_2$.
With the aid of $g_{ij}$ it is possible to define the thermodynamic curvature, $R=2 R_{1212}/g$
with $g=\mathrm{det}(g_{ij})$ and $R_{1212}$ corresponding to the only independent component of the Riemann's tensor
for a two-dimensional variety. As it is clear from the very definition of $g_{ij}$, the thermodynamic curvature
depends on the second and third order moments of the thermodynamic variables that are conjugated to $(\beta^1,\beta^2)$,
therefore it carries information about the fluctuation of the physical quantities in particular around a phase transition, where these
fluctuations are expected to be very large; for example, if $(\beta^1,\beta^2)=(1/T,-\mu/T)$ then $R$ contains information
about the fluctuations of energy and particle number.
One of the merits of the thermodynamic curvature is that in proximity of a second order phase transition
$|R|\propto\xi^d$ where $d$ denotes the spatial dimension and
$\xi$ is the correlation length: as a consequence, it is possible to grasp information about the correlation length
by means of thermodynamics only. This divergence is related to the vanishing of the determinant of the metric,
therefore the thermodynamic geometry gives information on the location of the phase transition in the $(\beta^i)$ space.

The main purpose of this article is to report on our study about the thermodynamic geometry,
and in particular on the thermodynamic curvature, of the Quark-Meson (QM) model of Quantum Chromodynamics (QCD),
see \cite{Ruggieri:2013cya,Ruggieri:2014bqa,Frasca:2011zn,Skokov:2010sf} and references therein.
It is well known that at zero baryon chemical potential,  QCD matter experiences a smooth crossover from
a low temperature confined phase in which chiral symmetry is spontaneously broken,
to a high temperature phase in which color is deconfined and chiral symmetry is approximately
restored \cite{Borsanyi:2013bia,Bazavov:2011nk,Cheng:2009zi,Borsanyi:2010cj,Borsanyi:2010bp}.
The situation is however unclear at finite baryon chemical potential for QCD with three colors, due to the
sign problem that forbids first principle calculations. Because of this, effective models have been used
to study the phase structure of QCD at finite $\mu$, and there is nowadays a consensus that regardless of the model used,
the smooth crossover becomes a first order phase transition if $\mu$ is large enough: this leads to speculate the existence
of a critical endpoint (CEP) in the $(T,\mu)$ plane at which the crossover becomes a true phase transition with divergent
susceptibilities, and this point marks the separation between the crossover on the one hand and the first order line
on the other hand. We consider here the QM model
at finite $T$ and $\mu$,  which has been applied many times to study the phase structure of QCD,
and we study its thermodynamic geometry
following the lines depicted in~\cite{Castorina:2019jzw} where a similar study has
been performed for the Nambu-Jona-Lasinio (NJL) model.
The advantage of using the QM model is its renormalizability, which removes the dependence of the results
on the effective ultraviolet cutoff that instead appears in NJL calculations.
Moreover, it is interesting to check how the predictions on the phase structure of QCD change
when different effective models are used: this not only can shed light on the qualitative picture,
but also put a quantitative statement on the theoretical uncertainty of model predictions,
for example on the location of the CEP.

We can anticipate here the main results. The curvature is found to be negative at low temperature,
as for an ideal fermion gas; then a change of sign is observed near the chiral crossover, where $R$
develops a local maximum which becomes more pronounced when the chemical potential is increased; finally,
$R$ becomes negative again at high temperature and approaches zero from below.
Moreover, the dependence of $R$ on temperature is nontrivial for large $\mu$ where two peaks are found in the crossover region,
one negative at smaller temperature and one positive at higher temperature.
Change of sign of $R$ has been observed for many substances
\cite{Ruppeiner:2008tcb,Ruppeiner:2012tct,Ruppeiner:2013tar,Ruppeiner:2013tfb,Ruppeiner:2015tsf,
Ruppeiner:2015tsw,Ruppeiner:2015trf,Ruppeiner:2016smt,Wei:2013ctb}
and it has been interpreted in terms of the nature of the attractive/repulsive microscopic interaction.
In the case of the QM model it is difficult to support the relation between the nature of the interaction
and the change of sign of $R$
since the interaction is always attractive:
although mathematically the change of sign can be understood in terms of the different sign that accompanies
the third order fluctuations in the expression of $R$,
it is not clear if the change of sign of $R$ has any physical meaning in the model at hand
and we think that this certainly deserves further study.
Moreover, the height of the peak of $R$ increases along the critical line as $\mu$ is increased from zero to the corresponding CEP value
and diverges at the CEP: this is in agreement with $|R|\propto\xi^3$ since the correlation
length remains finite at the crossover but increases as the crossover becomes sharper and eventually diverges at the critical endpoint.
We also discuss how the mixed susceptibility, $\langle\Delta U \Delta N\rangle$ which is nonzero at finite $\mu$,
is crucial to have $g=0$ at the CEP.

The plan of the article is as follows. In Section II we briefly review the thermodynamic geometry
and in particular the thermodynamic curvature. In Section III we review the QM model.
 In Section IV we discuss
$R$ for the QM model. Finally, in Section V we draw our conclusions.
We use the natural units system $\hbar=c=k_B=1$ throughout this article.

\section{Thermodynamic curvature}
The idea of thermodynamic fluctuations, thermodynamic geometry and in particular of thermodynamic curvature
is pretty old \cite{Weinhold:1975get,Weinhold:1975gtii} and is nowadays
introduced on several textbooks of Statistical Mechanics, see for example \cite{Ruppeiner:maybe,book:pathria,book:landau}.
We present here only the few concepts that are closely related to our work, while we refer to \cite{Ruppeiner:1995rgf,Ruppeiner:2010tci} and references
therein for more details.

Let us consider a thermodynamics system in the grand-canonical ensemble: we assume that its thermodynamic state at equilibrium
is specified in terms of  the coordinates $(T,\mu)$, where $T$ is the temperature and $\mu$ is the chemical potential
conjugated to the particle density; alternatively we can use a different set of coordinates, namely $(\beta,\gamma)$
where $\beta=1/T$ and $\gamma=-\mu/T$.
It is possible to build up a metric space in the $(\beta,\gamma)$ variety by defining a distance, namely
\begin{equation}
d\ell^2 = g_{\beta\beta}d\beta d\beta + 2 g_{\beta\gamma}d\beta d\gamma + g_{\gamma\gamma}d\gamma d\gamma,\label{Eq:distance_1}
\end{equation}
where for classical systems with grand-canonical partition function ${\cal Z}$ \cite{Janyszek:1989rtm}
\begin{equation}
g_{ij} \equiv \frac{\partial\log{\cal Z}}{\partial\beta^i \partial\beta^j}
=\frac{\partial\phi}{\partial\beta^i \partial\beta^j}\equiv\phi_{ij},\label{eq:tem}
\end{equation}
and $\phi\equiv\beta P$, $P=-\Omega$ with $\Omega$ representing the thermodynamic potential
per unit volume; moreover, $\beta^1 = 1/T$, $\beta^2=\gamma=-\mu/T$.
The line element $d\ell^2$ in Eq.~\eqref{Eq:distance_1} represents
effectively a distance in the 2-dimensional variety, in the sense that the probability to
fluctuate from the equilibrium state $S_1=(\beta,\gamma)$ to another equilibrium state $S_2=(\beta+d\beta,\gamma+d\gamma)$ is
\begin{equation}
\frac{dP}{d\beta d\gamma} \propto \sqrt{g} \exp\left(-\frac{d\ell^2}{2}\right);
\label{eq:probability}
\end{equation}
therefore, the larger the distance the less probable is to have a fluctuation from $S_1$ to $S_2$ and the two states are effectively distant.
In the above equation $g$ denotes the determinant of the metric tensor in Eq.~\eqref{eq:tem}.
With these definitions the thermodynamic curvature of the 2-dimensional variety is given by
\begin{equation}
R = -\frac{1}{2g^2}\left|
\begin{array}{ccc}
  \phi_{\beta\beta} &  \phi_{\beta\gamma} &  \phi_{\gamma\gamma} \\
  \phi_{\beta\beta\beta} &  \phi_{\beta\beta\gamma} &  \phi_{\beta\gamma\gamma} \\
 \phi_{\beta\beta\gamma} &  \phi_{\beta\gamma\gamma} &  \phi_{\gamma\gamma\gamma}
\end{array}
\right|,\label{eq:R_1}
\end{equation}
where $||$ means the determinant of the $3\times 3$ matrix and
\begin{eqnarray}
&&\phi_{ij} =   \langle(F_i - \langle F_i\rangle)(F_j - \langle F_j\rangle)\rangle,\\
&&\phi_{ijk}  \equiv \frac{\partial\phi_{jk}}{\partial\beta^i}= -\langle(F_i - \langle F_i\rangle)(F_j - \langle F_j\rangle)(F_k - \langle F_k\rangle)\rangle,\nonumber\\
&&
\end{eqnarray}
where $F_i$ denotes the physical quantity conjugated to $\beta^i$
and $\langle\rangle$ is the standard ensemble average.
We notice that our sign convention
agrees with \cite{Janyszek:1989rtm}, in particular
$R>0$ for the sphere. For our choice of coordinates we have, for example \cite{Janyszek:1},
\begin{eqnarray}
&&\phi_{\beta\beta} = \langle (U - \langle U\rangle)^2\rangle,\\
&&\phi_{\beta\gamma} = \langle (U - \langle U\rangle)(N - \langle N\rangle)\rangle,\\
&&\phi_{\gamma\gamma} = \langle (N - \langle N\rangle)^2\rangle,
\end{eqnarray}
where $U$, $N$ denote the internal energy and the particle number respectively.
The diagonal matrix elements $\phi_{\beta\beta}$ and $\phi_{\gamma\gamma}$, are related to the specific heat and the isothermal
compressibility, $\chi_T$, respectively~\cite{Janyszek:1}, namely
\begin{eqnarray}
&&\beta^2 \phi_{\beta\beta} = \left(\frac{\partial U}{\partial T}\right)_\gamma,\label{eq:Cherry1}\\
&&\beta \phi_{\gamma\gamma} = \frac{N^2}{V}\chi_T,\label{eq:Cherry2}
\end{eqnarray}
where $\chi_T = -(\partial V/\partial P)_{T}$.
Similarly, we can write
\begin{eqnarray}
&&\phi_{\beta\beta\beta} = -\langle (U - \langle U\rangle)^3\rangle,\\
&&\phi_{\gamma\gamma\gamma} = -\langle (N - \langle N\rangle)^3\rangle.
\end{eqnarray}

The thermodynamic curvature has the merit that close to a phase transition $|R|\propto\xi^3$ in three spatial dimensions,
where $\xi$ corresponds to the correlation length.  Therefore, in principle it is possible to access to microscopic
details like $\xi$ in the proximity of the phase transition just by means of thermodynamics.
Within our sign convention, $R<0$ for an ideal fermion gas and $R>0$ for an ideal boson gas;
$R=0$ for the ideal classical gas; for anyon gases \cite{Mirza:2008rag,Mirza:2009nta} it is possible to deform continuously the distribution
from a fermionic to a bosonic one, and this results in a change of the sign of $R$ in agreement with the
previous statement.
For interacting systems the interpretation of the sign of $R$
is far more complicated and nowadays there is no consensus on what this sign means. For example, for many substances it has been found that
$R<0$, but for these there exist a range of temperature/density in which $R>0$ and this change of sign
has been interpreted as a transition from a fluid to a solid-like fluid behavior \cite{Ruppeiner:2012tct,Ruppeiner:2015trf}.
In addition to this, it has been found that $R>0$ if the attractive interaction dominates,
while $R<0$ if the repulsive interaction is more important \cite{Ruppeiner:2013tar}: while this seems to be satisfied by several substances,
it is unclear if this relation between the sign of $R$ and the nature of the interaction is general.

\begin{widetext}
\section{The quark-meson model}
The QM model is an effective model of QCD in which quarks and mesons are considered on the same footing;
it is a very well known model in Quantum Field Theory, where it is called the linear-sigma model coupled to fermions
(see e.g. \cite{Peskin:1995ev,Weinberg:1996kr}).
 The meson part of the lagrangian density of the QM model is
\begin{equation}
{\cal L}_\mathrm{mesons} = \frac{1}{2}\left(
\partial^\mu\sigma\partial_\mu\sigma + \partial^\mu\bm\pi\cdot\partial_\mu\bm\pi
\right) - \frac{\lambda}{4}\left(\sigma^2 + \bm\pi^2 - v^2\right)^2 +h\sigma,\label{eq:ls1_aa}
\end{equation}
where $\bm\pi = (\pi_1,\pi_2,\pi_3)$ corresponds to the pion isotriplet field. This lagrangian density is
invariant under $O(4)$ rotations. On the other hand, as long as $v^2 > 0$ the potential
develops an infinite set of degenerate minima. We choose one ground state, namely
\begin{equation}
\langle\bm\pi\rangle=0,~~\langle\sigma\rangle=v=F_\pi,\label{eq:gs_aa}
\end{equation}
where $F_\pi\approx 93$ MeV denotes the pion decay constant. The ground state specified in Eq.~\eqref{eq:gs_aa}
breaks the $O(4)$ symmetry down to $O(3)$ since the vacuum is invariant only under the rotations of the pion fields.
The quark sector of the QM model is described by the lagrangian density
\begin{equation}
{\cal L}_\mathrm{quarks} = \bar\psi\left(
i\partial_\mu\gamma^\mu - G(\sigma +i\gamma_5 \bm\pi\cdot\bm\tau)
\right)\psi,\label{eq:qlg_aaa}
\end{equation}
where $\bm\tau$ are Pauli matrices in the flavor space. In the ground state~\eqref{eq:gs_aa} quarks get a dynamical (that is,
a constituent) mass given by
\begin{equation}
M_q = G\langle\sigma\rangle = G F_\pi.\label{eq:pppAAA}
\end{equation}
 We notice that in Eq.~\eqref{eq:qlg_aaa} there is no explicit mass term for the quarks. As a matter of fact,
in this effective model the explicit breaking of chiral symmetry   is achieved by
  the term
\begin{equation}
{\cal L}_\mathrm{mass} = h\sigma\label{eq:mass_term_aaa}
\end{equation}
in Eq.~\eqref{eq:ls1_aa}. In the limit $h/F_\pi^3\ll 1$ this implies $F_\pi M_\pi^2 = h$.
 Although in Eq.~\eqref{eq:qlg_aaa}  there is no explicit mass term, quarks get a constituent mass because of the spontaneous
breaking of the $O(4)$ symmetry in the meson sector: this implies that the quark chiral condensate can be nonzero.
Finally,
\begin{equation}
{\cal L}_\mathrm{QM} = {\cal L}_\mathrm{quarks} + {\cal L}_\mathrm{mesons}.
\end{equation}

The mean field effective potential of the QM model at zero temperature is given by
\begin{equation}
\Omega = U + \Omega_q,\label{eq:ep1}
\end{equation}
where
\begin{equation}
U = \frac{\lambda}{4}\left(\sigma^2 + \bm\pi^2 - v^2\right)^2 - h\sigma,\label{eq:ls1_aaMMM}
\end{equation}
is the classical potential of the meson fields as it can be read from Eq.~\eqref{eq:ls1_aa} and
\begin{equation}
\Omega_q = -2N_c N_f\int\frac{d^3p}{(2\pi)^3} E_p  \label{eq:ls1_aaMMMa}
\end{equation}
is the one-loop quark contribution, with
\begin{equation}
E_p = \sqrt{p^2 +M^2},~~~M=G\sigma.
\end{equation}
We notice that the quark mass depends on the field $\sigma$, thus
Eq.~\eqref{eq:ep1} is the effective potential for the $\sigma$ field computed at one-loop
and after regularization it corresponds to the condensation energy, namely the difference between
the energy of the state with $\langle\sigma\rangle\neq 0$ and $\langle\sigma\rangle= 0$ at $T=0$.

The quark loop in Eq.~\eqref{eq:ls1_aaMMMa} is divergent in the ultraviolet (UV) but the QM model is renormalizable,
therefore we can apply a strandard renormalization procedure to remove this divergence.
To this end we introduce the function \cite{Ruggieri:2013cya,Ruggieri:2014bqa,Frasca:2011zn}
\begin{equation}
\Omega_q(s) = -2N_c N_f\nu^{2s}\int\frac{d^3p}{(2\pi)^3} \left(p^2 +M^2\right)^{\frac{1}{2} -s},\label{eq:ls1_aaMMMabc}
\end{equation}
where $s$ is a complex number
 and $\nu$ carries the dimension of a mass in order to balance the wrong mass dimension of the integrand when $s\neq 0$.
The strategy is to compute the above integral for a finite value of $s$, then make analytical continuation to $s=0$.
The integral can be performed analytically with the result
\begin{equation}
\Omega_q(s) = - \frac{N_c N_f}{4\pi^{3/2}}\nu^{2s} M^{4-2s}\frac{\Gamma(s-2)}{\Gamma\left(s-\frac{1}{2}\right)},\label{eq:ThisExpr}
\end{equation}
where $\Gamma$ denotes the standard Euler's function. In the limit $s \rightarrow 0^+$ we get
\begin{equation}
\Omega_q(s) = - \frac{N_c N_f}{2\pi^{2}}\left[
-\frac{M^4}{8s} + \frac{M^4}{16}\left(
-3 +2\gamma_E +2\psi(-1/2) + 4\log\frac{M}{\nu}
\right)
\right],\label{eq:ThisExpr2}
\end{equation}
where $\psi(z)=d\log\Gamma(z)/dz$ and $\gamma_E \approx 0.577$ is the Euler constant.
We notice that the  UV divergence is now manifest in the analytical structure
of $\Omega_q(s)$ in the complex plane, namely it appears as a simple pole at $s=0$.

We now add two counterterms,
\begin{equation}
\Omega_\mathrm{ct} = \frac{\delta v}{2}\sigma^2 + \frac{\delta\lambda}{4}\sigma^4, \label{eq:divergence_2}
\end{equation}
to the quark loop and require that the renormalized quark loop does not shift the values of the $\sigma$ meson mass
as well as the value of the condensate obtained within the classical potential~\eqref{eq:gs_aa};
$\delta v$ and $\delta\lambda$ will absorb the divergence of $\Omega_q$ and the final result will be a convergent quantity.
By definining the renormalized potential as
\begin{equation}
\Omega_q^\mathrm{ren} = \Omega_q + \Omega_\mathrm{ct} \label{eq:divergence_3}
\end{equation}
the renormalization conditions thus read
\begin{equation}
\left.\frac{\partial\Omega_q^\mathrm{ren}}{\partial\sigma}\right|_{\sigma= v} = 0,~~
\left.\frac{\partial^2\Omega_q^\mathrm{ren}}{\partial\sigma^2}\right|_{\sigma= v} = 0;\label{eq:divergence_4}
\end{equation}
as a matter of fact, the first condition implies that $\Omega_q^\mathrm{ren}$ does not shift the global minimum of the
effective potential, and this will be given by the minimum of the classical potential, while the second condition
implies that for $\sigma=v$, namely in the ground state, the mass matrix of the $\sigma$ meson is not affected by the
quark loop.
The counterterms can be computed very easily and their expression is not necessary here, therefore we write directly
the expression for the renormalized potential, namely
\begin{equation}
\Omega_q^\mathrm{ren} = \frac{3G^4}{32\pi^2}N_c N_f \sigma^4
 - \frac{G^4 F_\pi^2}{8\pi^2} N_c N_f \sigma^2 + \frac{G^4 N_c N_f}{8\pi^2}\sigma^4\log\frac{F_\pi}{\sigma}. \label{eq:divergence_5aaa}
\end{equation}
We notice that $\Omega_q^\mathrm{ren}=0$ for $\sigma=0$ and that
$\Omega_q^\mathrm{ren}=-G^4 F_\pi^4 N_c N_f/32\pi^2$ for $\sigma=F_\pi$:
thus, $\Omega_q^\mathrm{ren}$ lowers the energy difference between the states with and without a condensate.

The finite temperature thermodynamic potential is finite and does not need any particular treatment:
it is given by the standard relativistic fermion gas contribution, namely
\begin{equation}
\Omega_T = -2N_c N_f T \int\frac{d^3p}{(2\pi)^3}\log\left(1 + e^{-\beta(E_p-\mu)}\right)\left(1 + e^{-\beta(E_p+\mu)}\right),
\end{equation}
where $\mu$ corresponds to the chemical potential and $E_p = \sqrt{p^2 +M^2}$
with $M=g \sigma$. Putting all together we get
\begin{eqnarray}
\Omega &=& U(\sigma) +\frac{3G^4}{32\pi^2}N_c N_f \sigma^4
 - \frac{G^4 F_\pi^2}{8\pi^2} N_c N_f \sigma^2 + \frac{G^4 N_c N_f}{8\pi^2}\sigma^4\log\frac{F_\pi}{\sigma}\nonumber\\
 &&-2N_c N_f T \int\frac{d^3p}{(2\pi)^3}\log\left(1 + e^{-\beta(E_p-\mu)}\right)\left(1 + e^{-\beta(E_p+\mu)}\right).
\end{eqnarray}
The pressure is $P(T,\mu)=-\Omega(T,\mu,\bar\sigma)$ where $\bar\sigma$ is the value of $\sigma$ that minimizes $\Omega$
at a given $(T,\mu)$.

\end{widetext}

\section{Results}

\subsection{The thermodynamic curvature}
\begin{figure}[t!]
\begin{center}
\includegraphics[scale=0.15]{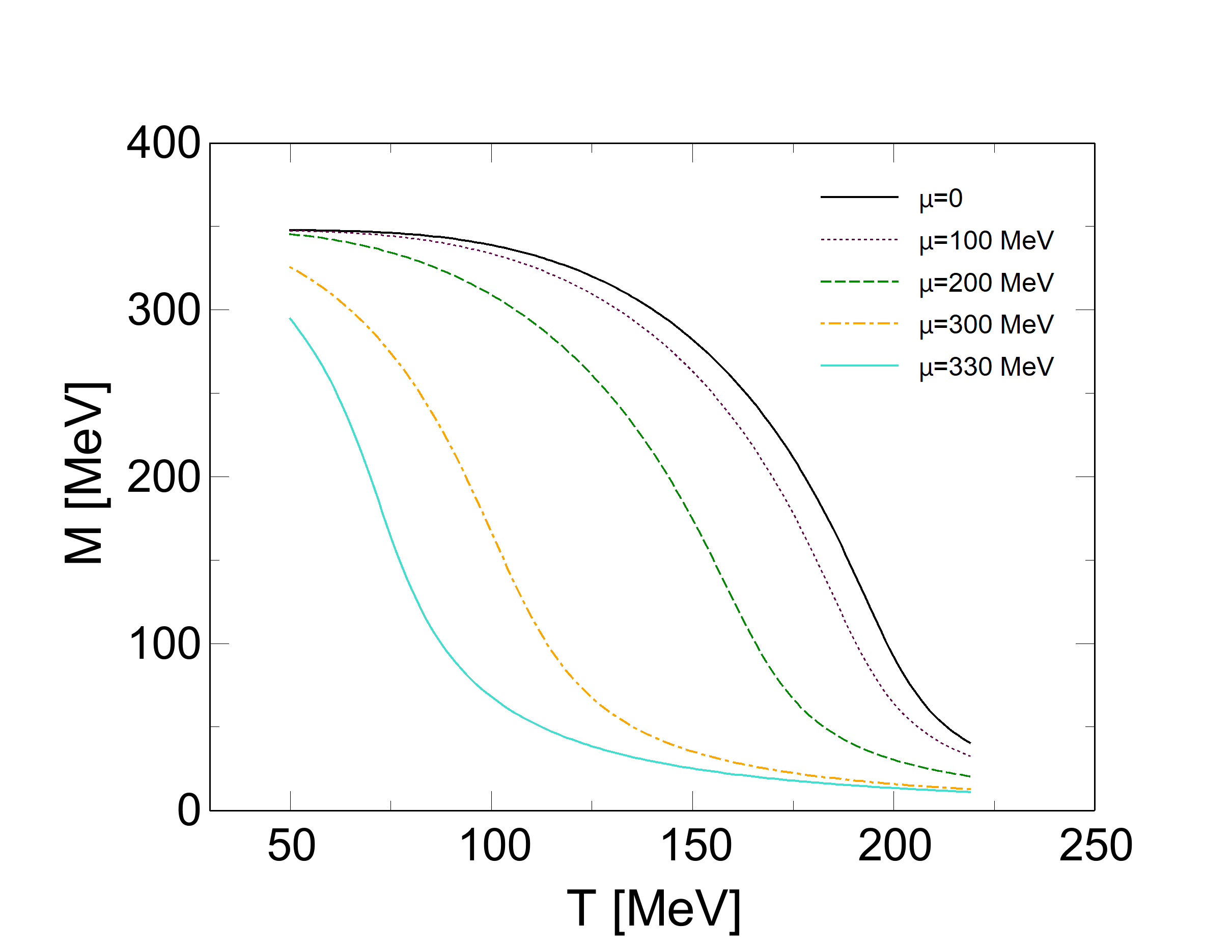}
\end{center}
\caption{\label{Fig:arianna_0}Constituent quark mass, $M$, versus temperature
for several values of the quark chemical potential:
black solid line is for $\mu=0$, brown dotted line denotes $\mu=100$ MeV, green dashed line stands for $\mu=200$ MeV,
orange dot-dashed line corresponds to $\mu=300$ MeV and finally
turquoise solid line stands for $\mu=330$ MeV.
}
\end{figure}

In this section we summarize the results we have obtained for the QM model.
Our parameter set is $M_\sigma=700$ MeV, $v=F_\pi=93$ MeV, $M_\pi=138$ MeV and $G=3.6$; these give
$M=335$ MeV at $T=0$ and $\mu=0$, $\lambda=M_\sigma^2/2F_\pi^2=28.3$ and $h=M_\pi^2 F_\pi=1.78\times 10^6$ MeV$^3$.
We have also used another parameter set with $M_\sigma=600$ MeV and $M=350$ MeV at $\mu=T=0$ but the results are unchanged qualitatively,
therefore we present here only the results related to the first parameter set.

In Fig.~\ref{Fig:arianna_0} we plot the constituent quark mass as a function of temperature for several values
of the quark chemical potential: black solid line is for $\mu=0$, brown dotted line denotes $\mu=100$ MeV, green dashed line stands for $\mu=200$ MeV,
orange dot-dashed line corresponds to $\mu=300$ MeV
and finally turquoise solid line stands for $\mu=330$ MeV.
For any value of $\mu$ there exists a range of temperature in which $M$ decreases: this is the chiral crossover
from a low temperature phase with spontaneous chiral symmetry breaking from a high temperature phase
in which chiral symmetry is approximately restored. The larger $\mu$ the sharper the change of $M$ with $T$ is,
and for $(T,\mu)\equiv(T_E,\mu_E)\approx(30~\mathrm{MeV},360~\mathrm{MeV})$ we find the critical endpoint (CEP) at which the crossover
becomes a true second order phase transition and for $\mu>\mu_E$ the phase transition is a first order one.

\begin{figure}[t!]
\begin{center}
\includegraphics[scale=0.15]{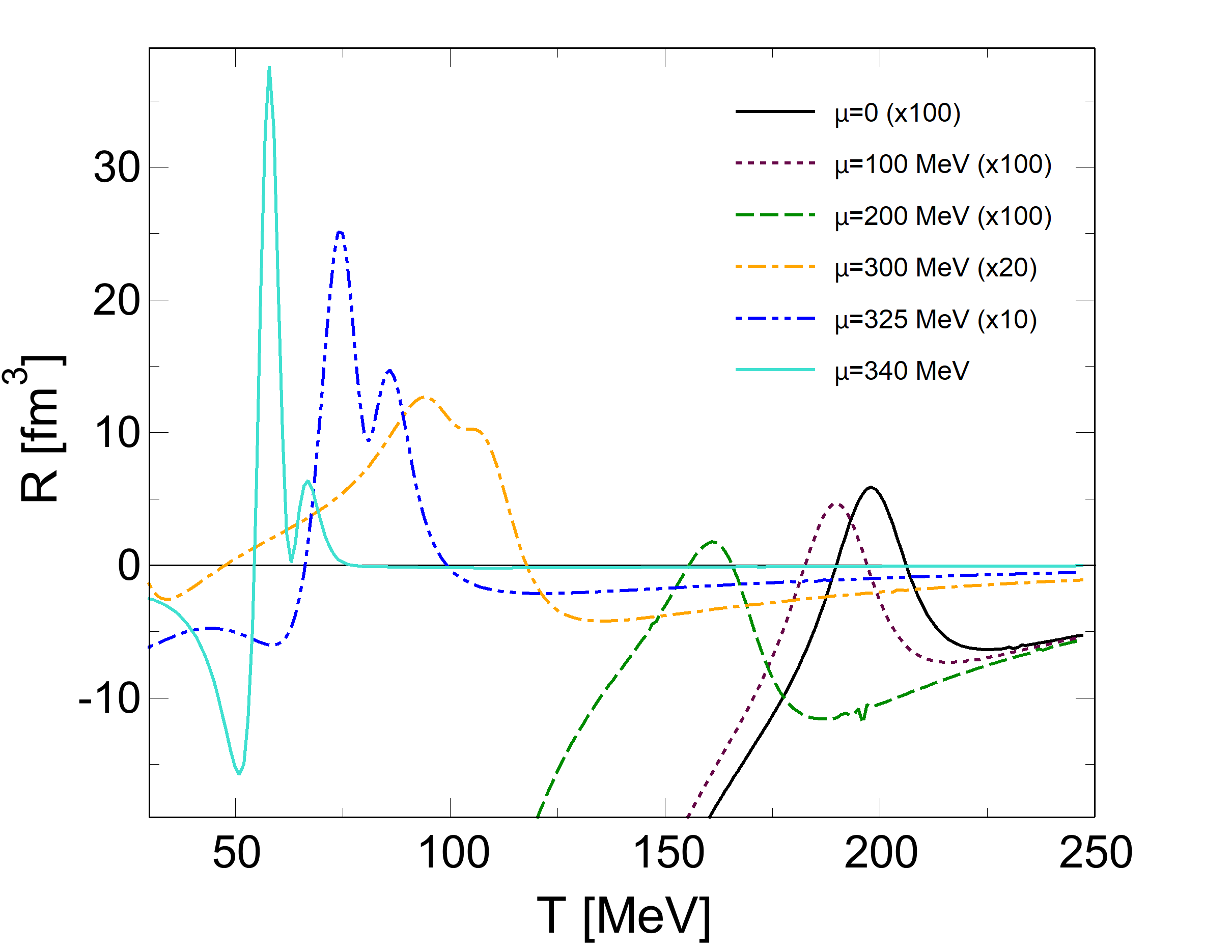}
\end{center}
\caption{\label{Fig:arianna_1}Thermodynamic curvature, $R$, versus temperature
for several values of the quark chemical potential:
black solid line is for $\mu=0$, brown dotted line denotes $\mu=100$ MeV, green dashed line stands for $\mu=200$ MeV,
orange dot-dashed line corresponds to $\mu=300$ MeV, blue dot-dot-dashed line denotes $\mu=325$ MeV and finally
turquoise solid line stands for $\mu=340$ MeV.
}
\end{figure}

In Fig.~\ref{Fig:arianna_1} we plot the thermodynamic curvature, $R$, versus temperature
for several values of the quark chemical potential:
black solid line is for $\mu=0$, brown dotted line denotes $\mu=100$ MeV, green dashed line stands for $\mu=200$ MeV,
orange dot-dashed line corresponds to $\mu=300$ MeV, blue dot-dot-dashed line denotes $\mu=325$ MeV and finally
turquoise solid line stands for $\mu=340$ MeV.
At small temperature the curvature is negative, as  for a free fermion gas.
However, we notice that the sign of $R$ changes around the crossover, then becoming negative
again for $T\gg T_c$: the crossover corresponds to a change in the geometry
from hyperbolic to elliptic \cite{Saccheri,Beltrami:1828,book:1,Lobachevsky:1829}.
Following \cite{Castorina:2019jzw} we identify the region in which $R>0$ with the crossover.
This interpretation is supported by the fact that the local maxima of $R$ appear to be very close to those of $|dM/dT|$,
the latter giving a rough location of the crossover itself, see also below.
The fact that the magnitude of $R$
in the critical region remains small for small $\mu$ is related to the fact that in this region the crossover is very smooth;
on the other hand, when we approach the critical endpoint the crossover is closer to a second order
phase transition and $R$ develops clear peaks. 

We also notice that the structure of $R(T)$ as the critical endpoint is approached is quite interesting.
Indeed, for $\mu=340$ MeV in Fig.~\ref{Fig:arianna_1} we find that $R$ is negative and drops down before rising to a positive peak around the crossover.
The behavior of $R$ that we find can be understood mathematically
since $R$ combines several second and third order cumulants
with different signs, see Section II.
Overall, the increase of the magnitude of $R$ as the CEP is approached is due to the determinant of the metric that becomes small around CEP and
eventually vanishes at the CEP, see below.
We also notice that increasing the temperature right above the peak results in $R=0$ then $R$ stays positive for a substantial
temperature range, before becoming negative again: the $R=0$ point can be understood since
$\phi_{\beta\beta\beta}$, $\phi_{\beta\beta\gamma}$, $\phi_{\beta\gamma\gamma}$ and
$\phi_{\gamma\gamma\gamma}$ vanish at that temperature and so $R$ does:
this is in agreement with well known fact that the third order cumulants change sign
around the critical endpoint \cite{Vovchenko:2015pya,Li:2017ple,Li:2018ygx,Shao:2017yzv}

\subsection{The thermodynamic geometry at the critical line}

\begin{figure}[t!]
\begin{center}
\includegraphics[scale=0.15]{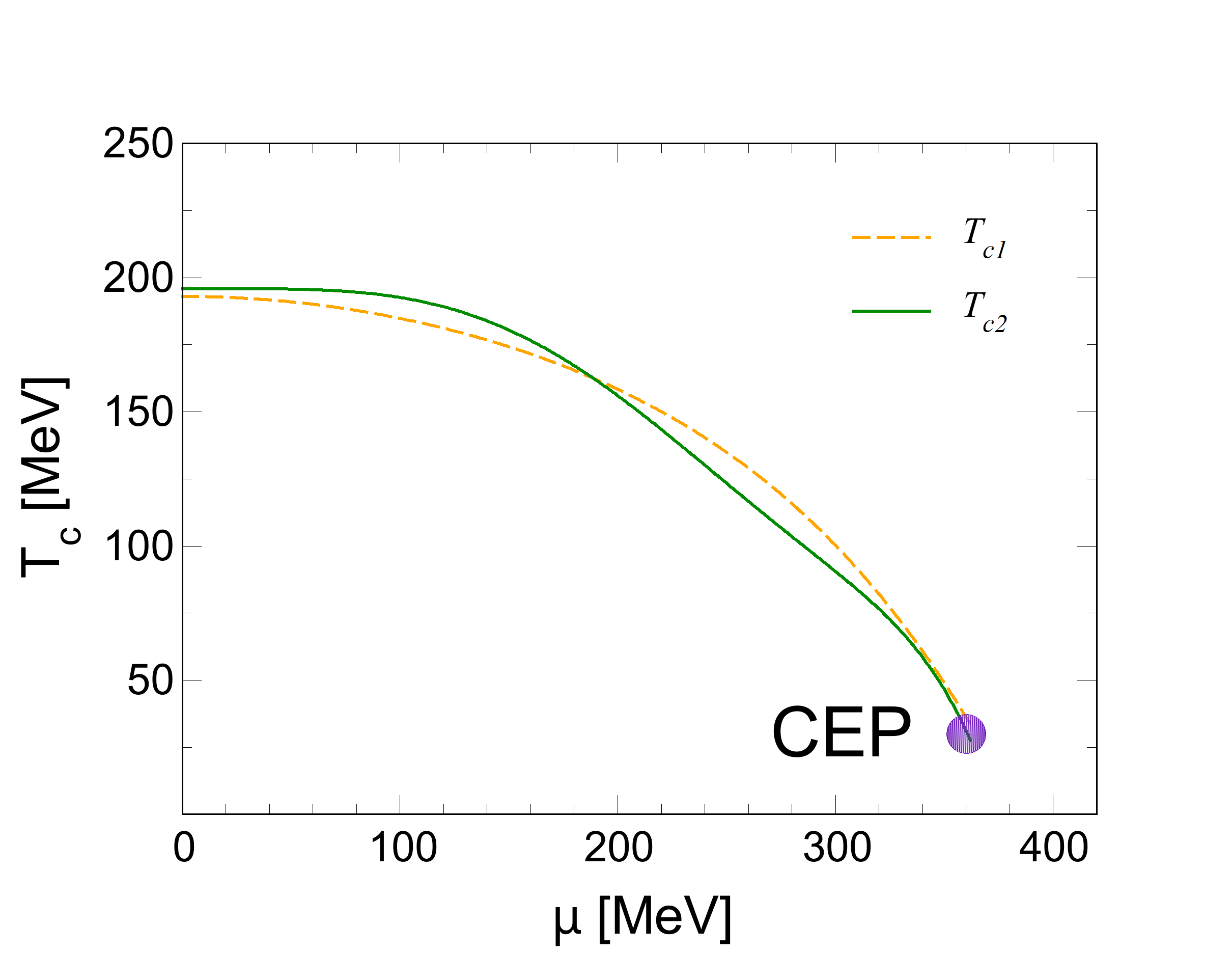}
\end{center}
\caption{\label{Fig:arianna_0m1}  Critical temperatures, $T_{c1}$ and $T_{c2}$, as a function of the chemical potential.
The indigo dot denotes the critical end point (CEP).
}
\end{figure}

In the model at hand, as well as in full QCD, there is no real phase transition at high temperature and small chemical potential, rather only a smooth crossover.
Because of this, the location of the critical temperature is ambiguous: for example, the crossover region can be identified around
the temperature at which $|dM/dT|$ is maximum, or by the location of the peak of the chiral susceptibility.
We define two critical temperatures:
\begin{eqnarray}
&&T_{c1}:~\mathrm{from~the~peak~of~}\left|\frac{dM}{dT}\right|,\\
&&T_{c2}:~\mathrm{from~the~positive~peak~of~}R .
\end{eqnarray}
In particular, using $T_{c1}$ we define the crossover at a given $\mu$ by choosing the temperature at which the constituent quark mass
has its maximum change.
In Fig.~\ref{Fig:arianna_0m1}
we plot $T_{c1}$ and $T_{c2}$ as a function of the chemical potential. The two lines end up and coincide at the critical endpoint,
which is denoted by an indigo dot.
The two critical temperatures differ for few percent at most, therefore the local maxima of $R$ in the $(T,\mu)$ plane are very close
to the points at which the constituent quark mass has its maximum change
which supports the idea that the peaks of $R$ do relate to the chiral crossover.

\begin{figure}[t!]
\begin{center}
\includegraphics[scale=0.15]{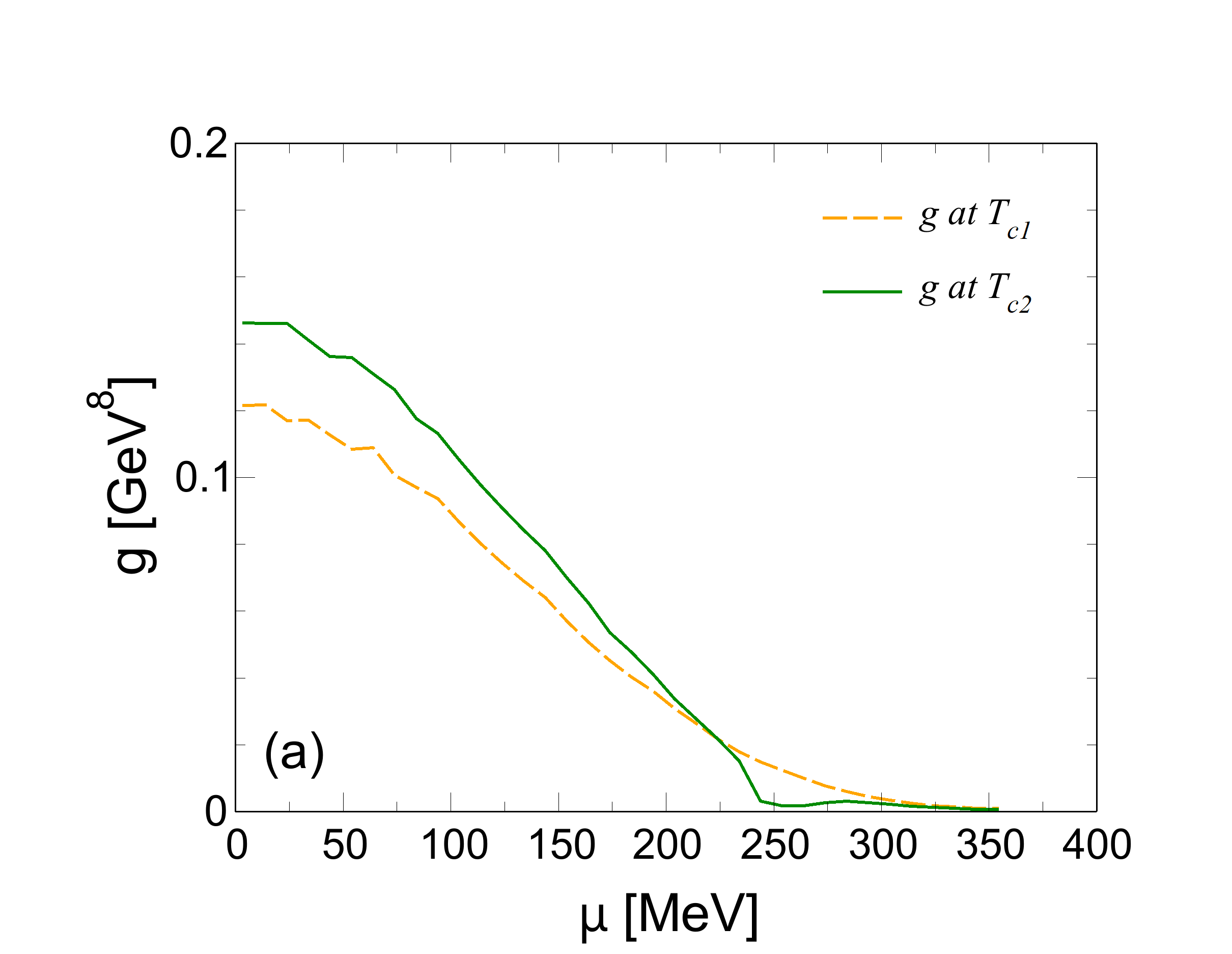}\\
\includegraphics[scale=0.15]{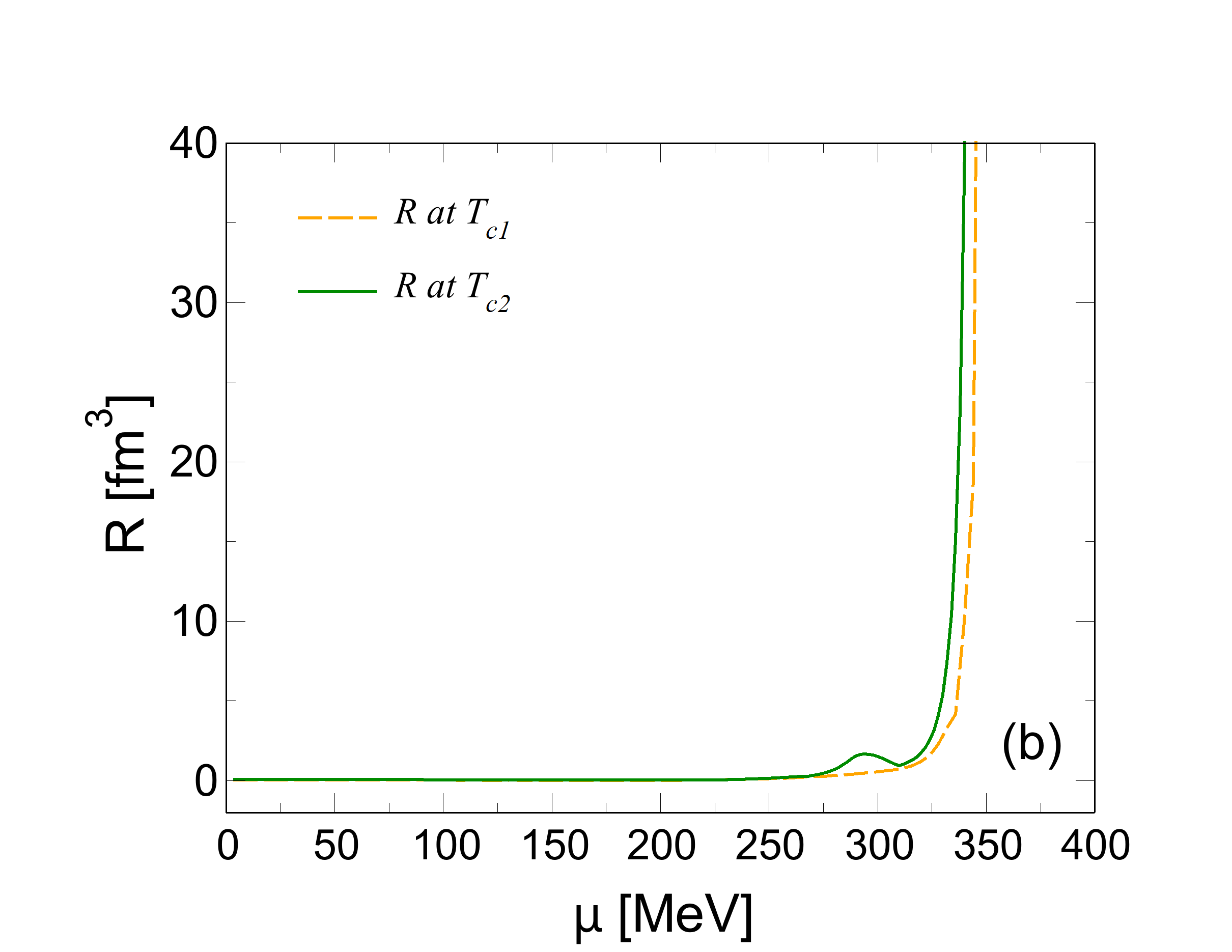}
\end{center}
\caption{\label{Fig:arianna_01} Panel {\bf (a)}. Determinant of the metric:
orange dashed line corresponds to the value computed $T_{c1}$, solid green line to the value computed at $T_{c2}$.
Panel {\bf (b)}. Thermodynamic curvature versus temperature. Color and line conventions are the same used
for Panel {\bf (a)}.
}
\end{figure}

In the panel {\bf (a)} of Fig.~\ref{Fig:arianna_01} we plot the determinant of the metric, $g$, in proximity of the critical line.
The orange dashed line corresponds to the value of $g$ computed at $T_{c1}$, while the solid green line denote the values of $g$ computed at $T_{c2}$.
We notice that $g$ is always positive in the crossover region hence thermodynamic distance
is well defined there and the system is thermodynamically stable. 
The mismatch between the two curves is clearly related to the definition used for the critical temperature;
nevertheless, the qualitative behavior of $g$ is the same in the two cases.
We also find that around the CEP the determinant is very small and eventually vanishes at the CEP,
as anticipated: the vanishing of the dererminant at the CEP is expected at a second order
phase transition on the base of thermodynamic stability \cite{CALLEN};
moreover, because $g=0$ at the CEP we get that $R$ diverges there, 
as it happens for example for the van der Waals gas \cite{geometrical:aspects,math_santoro,Janyszek:2}.

In the panel {\bf (b)} of Fig.~\ref{Fig:arianna_01} we plot $R$ at the critical temperature. Again, we compare
the result obtained using two different definitions of the critical temperature: the orange dashed line
denotes $R$ computed at $T_{c1}$, while the solid green line denote the values of $R$ computed at $T_{c2}$.
We notice that in both cases the qualitative behavior of $R$ is the same.
In particular, the magnitude of $R$ increases when $(T,\mu)$ approach the CEP and diverges at the CEP, in agreement with the
previous discussion. 
The divergence of $R$ at the CEP supports the idea
that $|R|$ measures the correlation volume around
the phase transition since the latter also diverges at the CEP \cite{Ruppeiner:1979trg}.

\begin{figure}[t!]
\begin{center}
\includegraphics[scale=0.15]{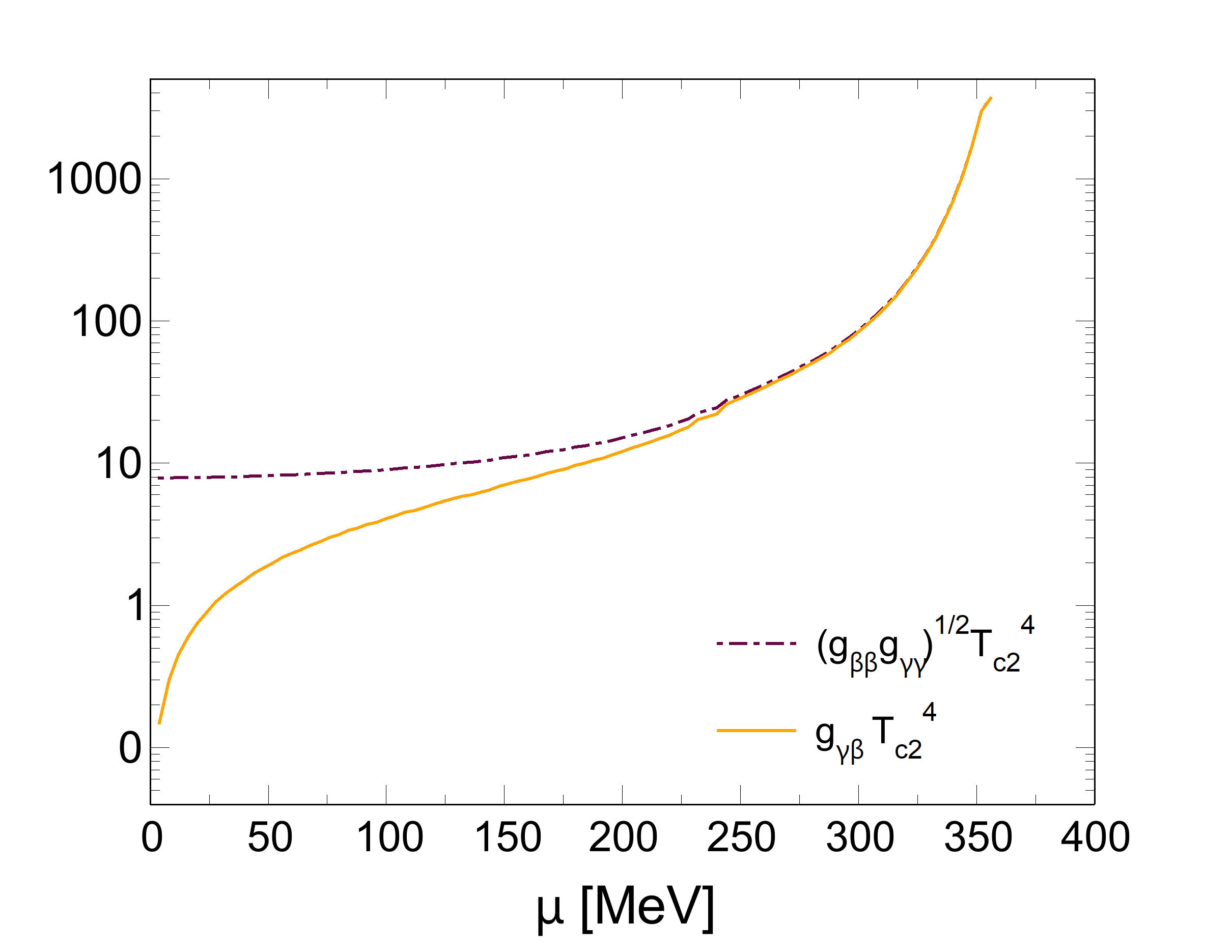}
\end{center}
\caption{\label{Fig:eliana_01} $T_{c2}^4 (g_{\beta\beta}g_{\gamma\gamma})^{1/2}$ (indigo dot-dashed line)
and $T_{c2}^4 g_{\beta\gamma}$ (orange solid line) as a function of the chemical potential along $T_{c2}$.
}
\end{figure}

The thermodynamic curvature diverges at the CEP because the determinant of the metric is zero there:
the condition $g=0$ corresponds to thermodynamic instability and thus to a phase transition. 
Clearly we can write (see Section II for more details)
\begin{eqnarray}
g &=& g_{\beta\beta}g_{\gamma\gamma} - g_{\beta\gamma}^2 \\
&=&\left\langle(\Delta U)^2\right\rangle\left\langle(\Delta N)^2\right\rangle
-\left\langle\Delta U\Delta N\right\rangle^2,\label{eq:mixed_1}
\end{eqnarray}
where in particular $\left\langle \Delta U\Delta N \right\rangle $ corresponds to the mixed energy-baryon number fluctuation.
In Fig.~\ref{Fig:eliana_01} we plot $T_{c2}^4 (g_{\beta\beta}g_{\gamma\gamma})^{1/2}$ (indigo dot-dashed line)
and $T_{c2}^4 g_{\beta\gamma}$ (orange solid line) as a function of the chemical potential along $T_{c2}$
(for $T_{c1}$ we get similar results therefore we do not show them here). At $\mu=0$ we find $g_{\beta\gamma}=0$
thus $g>0$; as $\mu$ is increased the mixed susceptibility rapidly grows up and hits eventually
$(g_{\beta\beta}g_{\gamma\gamma})^{1/2}$ leading to $g=0$ and to the divergent curvature.
We conclude that the CEP (i.e. the divergent curvature) occurs in the phase diagram because the underlying microscopic interaction
leads to a rapidly increasing mixed energy and baryon number fluctuation.
We notice that the vanishing of $g$ is something more than getting a divergent baryon number susceptibility at the CEP:
in fact, at the CEP all the matrix elements of the metric diverge, but it is the vanishing of the determinant
that guarantees that $R$ diverges and thus the crossover becomes a second order phase transition.
We will discuss how the mixed susceptibility is sensitive to the location of the CEP in a forthcoming article.

\begin{figure}[t!]
\begin{center}
\includegraphics[scale=0.12]{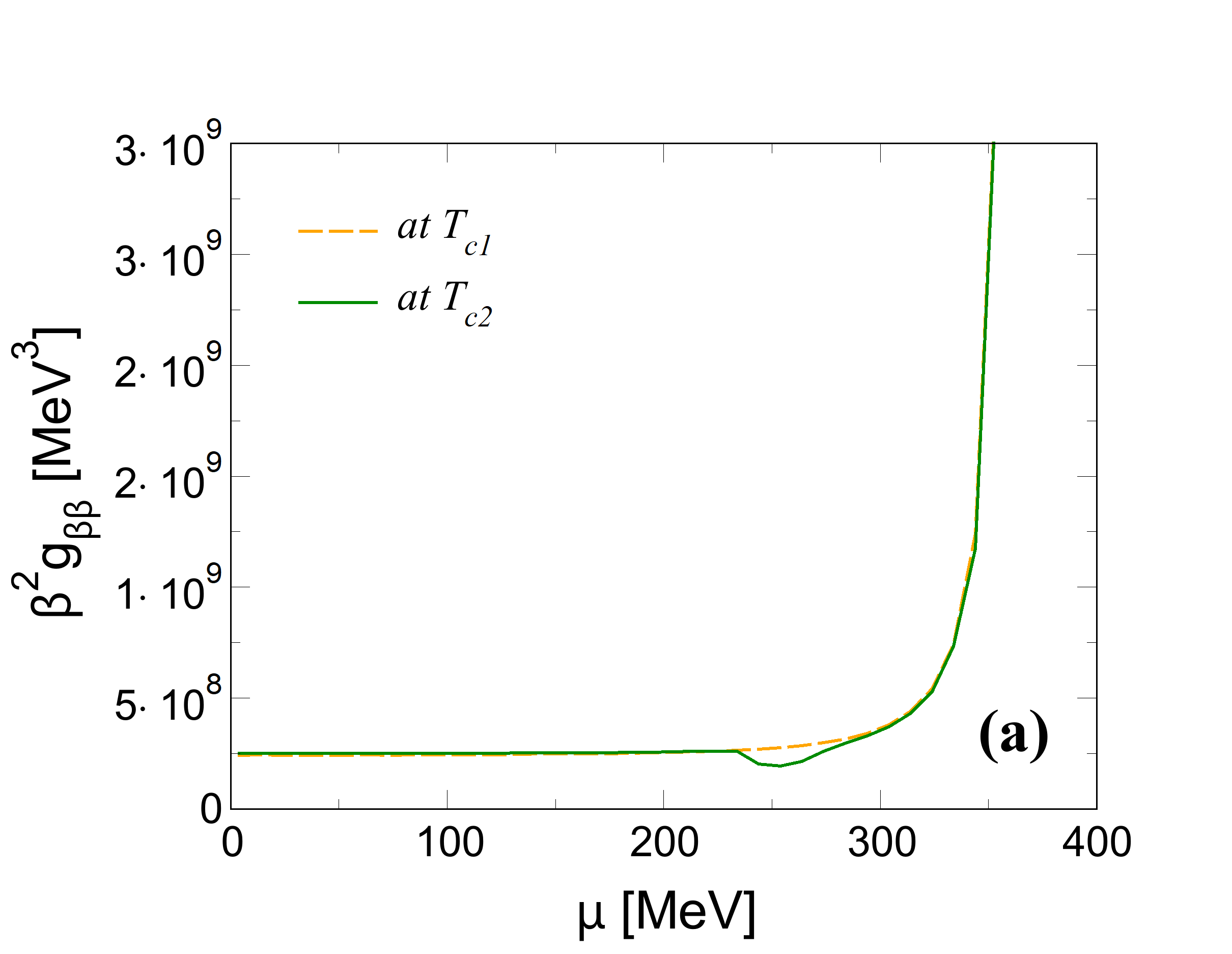}\\
\includegraphics[scale=0.12]{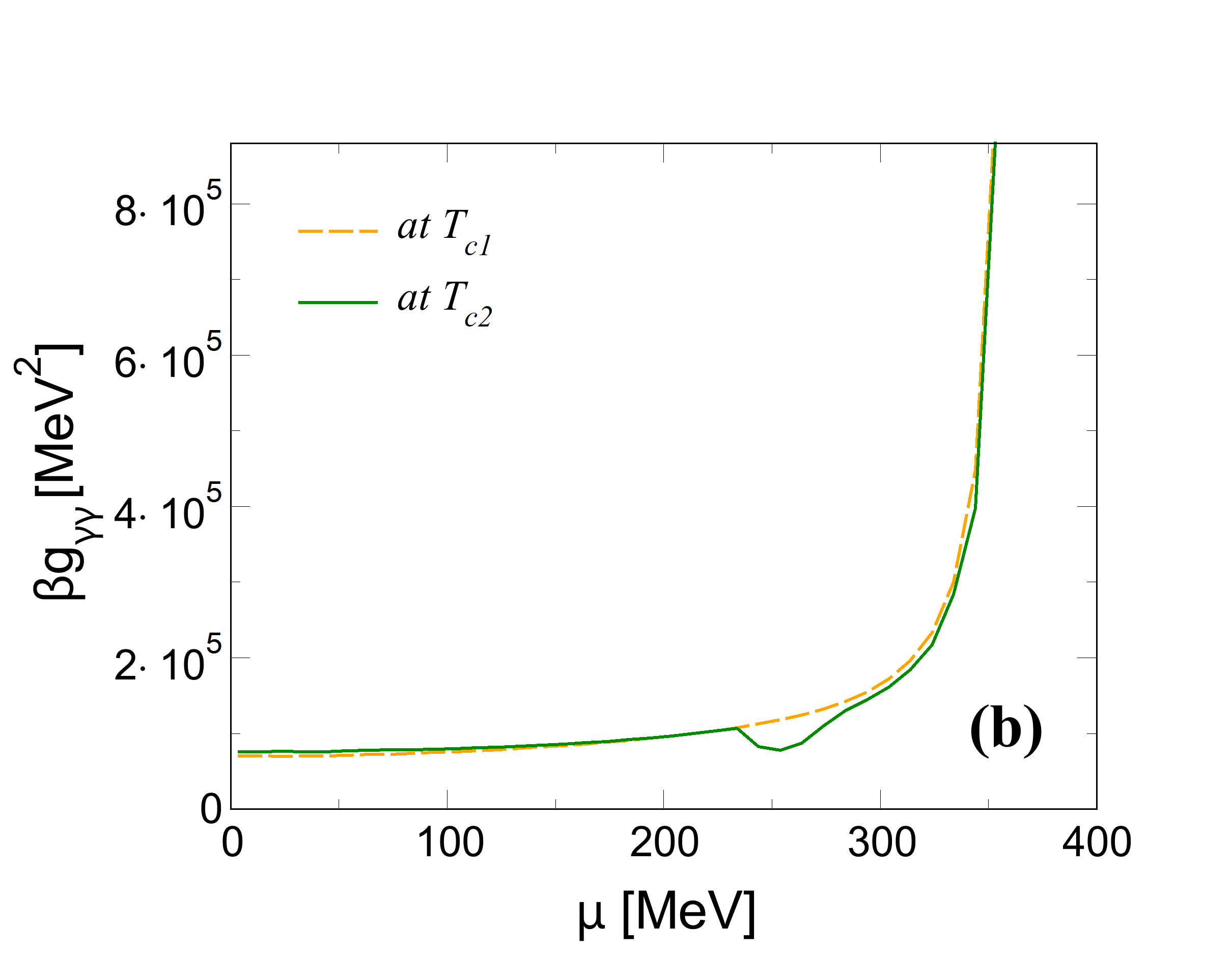}
\end{center}
\caption{\label{Fig:arianna_3} $\beta^2 g_{\beta\beta}$ (panel (a)), $\beta g_{\gamma\gamma}$ (panel (b))
at the critical line. Orange dashed line corresponds to data computed at $T_{c1}$ while green solid line denote data computed at $T_{c2}$.
}
\end{figure}

In Fig.~\ref{Fig:arianna_3} we plot
$\beta^2 g_{\beta\beta}$ (panel (a)) and $\beta g_{\gamma\gamma}$ (panel (b))
at the critical line. Orange dashed line corresponds to data computed at $T_{c1}$ while green solid line denote data computed at $T_{c2}$.
According to Eqs.~\eqref{eq:Cherry1} and~\eqref{eq:Cherry2}, $\beta^2 g_{\beta\beta}$ and $\beta g_{\gamma\gamma}$
are proportional to the specific heat and the isothermal compressibility respectively.
We notice that both quantities stay finite around the crossover at small $\mu$ but diverge as the critical endpoint is approached,
in agreement with the fact that crossover becomes a second order phase transition there.

\section{Summary and Conclusions}
We have applied the concept of thermodynamic geometry, in particular of thermodynamic curvature $R$,
to the chiral crossover of Quantum Chromodynamics (QCD) at finite temperature, $T$
and finite baryon chemical potential, $\mu$. The crossover has been modeled
by the renormalized Quark-Meson model (QM model) which is capable to describe the spontaneous breaking chiral symmetry.
Although thermodynamic geometry has been introduced many years ago, its use for the high temperature phase of QCD
has been only marginal. One of the merits of $R$ is that $|R|\propto \xi^3$ near a second order phase transition in three spatial dimensions,
where $\xi$ corresponds to the correlation length; in QCD a crossover is expected at high temperature instead of a real
second order phase transition therefore the interpretation of $R$ has to be done carefully,
nevertheless it is fair to relate the peaks of $R$ to the chiral crossover. We support this idea here, albeit some detail that in our opinion
deserve further studying.

We have studied $R$ for the QM model at finite $T$ and $\mu$. In the QM model the mass constituent quark mass, $M=M(T,\mu)$, is
related to the quark condensate and it is computed self-consistently.
We have found that for small values of $\mu$, where the model presents a very smooth crossover at finite temperature,
increasing temperature results in a change of sign from negative to positive in the crossover region,
as well as to a modest peak of $R$: this is similar to what has been observed in the
Nambu-Jona-Lasinio model \cite{Castorina:2019jzw}, and this peak appears in correspondence of the peak of $|dM/dT|$
so it is natural to identify the peak of $R$ with the chiral crossover.
The change of sign of $R$ has been interpreted previously as an indicator of the attractive/repulsive nature of the microscopic
interaction. However, in our model it is not clear whether this interpretation is legit because the interaction
is always attractive, despite this $R$ changes sign around the chiral crossover.
We think that this point deserves more attention and future studies.
We have also studied several matrix elements of the metric, which are related to the isothermal compressibility and to the specific heat,
as well as the curvature, at the critical line $T_c(\mu_c)$, finding the divergence of these as the critical endpoint is approached.
Overall, these results support the idea that although in QCD at small $\mu$ there is a smooth crossover rather than a phase transition,
the thermodynamic curvature is capable to capture this crossover by developing local maxima around $T_c$.

We have also pointed out that due to fluctuations of both energy and baryon number in the
grandcanonical ensemble, a mixed susceptibility, $\langle\Delta U\Delta N\rangle$,
develops at finite $\mu$. We have shown that
the CEP in the temperature and baryon chemical potential plane  occurs when the determinant of the themodynamic metric vanishes: this happens
$\langle\Delta U\Delta N\rangle$ grows up considerably at finite baryon chemical potential.

There are several aspects that deserve further investigations.
Firstly, it is interesting to study how the repulsive vector interaction affects $R$ at finite $T$ and $\mu$:
this indeed might shed some light on the connection between the attractive/repulsive nature of the interaction
and the sign of $R$.
To this end, it is useful to remark that effective models of QCD like the one used in this article
are ideal tools to study the thermodynamic curvature, for at least two reasons.
The first one is that the interaction is strong, so they allow to study quantitatively $R$ for systems
that are quite far from the ideal gas or the weakly coupled system: in fact,
most of the works done on $R$ in the last $\approx 40$ years hardly consider
strongly coupled systems, therefore a fundamental understanding of $R$ for these thermodynamic
systems is lacking.
In addition to this, the microscopic interaction in these models is under control, so it is possible to study
how the microscopic details affect $R$ around the phase transition.
Moreover, it is of a certain interest to study $R(T,\mu)$ around the QCD chiral crossover
using a Ginzburg-Landau effective potential, since this might lead to analytical expressions of the curvature
and help to prove quantitatively the relation $|R|\propto\xi^3$ for the model at hand.
Even more, it is certainly interesting to study the behavior of $R$ for higher dimensional varieties, for example
enlarging the present two-dimensional space by a third direction representing isospin or magnetic field.
Finally, it is well known that the QCD phase structure at large density is pretty rich: it is interesting to apply
the ideas of thermodynamic curvature in this regime as well.
All these interesting themes might be not shed new light on the phase structure of QCD, but we are confident that
they will help to understand more about the significance of the thermodynamic geometry.

Moreover, the thermodynamic geometry allows for a natural definition of the CEP since this can be identified
with the point in the $(T,\mu)$ plane where the determinant of the metric vanishes,
which is equivalent to a precise relation between susceptibilities at the CEP: it will be interesting to study
how the the susceptibilities at small and moderate $\mu$ are sensitive to the location of the CEP,
hopefully to get information that can be tested in first principle calculations and shed light on the
CEP in full QCD. We plan to report on this in a forthcoming article.

\section*{ACKNOWLEDGEMENTS}
The authors acknowledge Paolo Castorina, Daniele Lanteri, John Petrucci and Sijiang Yang for inspiration,
discussions and comments on the first version of this article.
The work of M. R. is supported by the National Science Foundation of China (Grants No.11805087 and No. 11875153)
and by the Fundamental Research Funds for the Central Universities (grant number 862946).

\end{document}